\documentclass[final,3p,times]{elsarticle}

\usepackage{amssymb}
\usepackage{amsmath}
\usepackage{xfrac}
\usepackage{enumerate}
\usepackage{lineno}
\usepackage{float}
\modulolinenumbers[5] 
\usepackage[hidelinks]{hyperref}
\usepackage[english]{babel}
\usepackage{microtype}
\usepackage{subcaption}
\usepackage{caption}


\journal{Nuclear Instruments and Methods in Physics Research A}

\begin{document}

\begin{frontmatter}

\title{\boldmath Options of RICH detectors based on silica aerogels for high momenta range}






\author[binpaddress,nstuaddress]{A.Yu. Barnyakov,\corref{cor1}}
\cortext[cor1]{Corresponding author.}
\ead{A.Yu.Baryakov@inp.nsk.su}
\author[binpaddress,nsuaddress]{V.S.Bobrovnikov}
\author[binpaddress]{A.R.Buzykaev}
\author[binpaddress,nstuaddress]{A.V. Chepelev,}
\author[binpaddress,nstuaddress]{R.A. Efremov,}
\author[bicaddress]{A.F. Daniluyk,}
\author[binpaddress,nsuaddress]{A.A. Katcin,}
\author[nsuaddress,binpaddress]{E.A.Kravchenko}
\author[binpaddress,nstuaddress]{I.A. Kuyanov}
\author[binpaddress,nstuaddress]{A.D. Ofitserov,}
\author[binpaddress,nsuaddress]{I.V.Ovtin}
\address[binpaddress]{Budker Institute of Nuclear Physics,  Acad.~Lavrentiev Prospect 11, 630090, Novosibirsk, Russia}
\address[nstuaddress]{Novosibirsk State Technical University, Karl~Marks Prospect 20, 630073, Novosibirsk, Russia}
\address[nsuaddress]{Novosibirsk State University, St.~Pirogova 1, 630090, Novosibirsk, Russia}
\address[bicaddress]{Boreskov Institute of Catalysis, Acad.~Lavrentiev Prospect 5, 630090, Novosibirsk, Russia} 


\begin{abstract}
Nowadays, several projects of future colliding beam experiments are being considered in the world. Among them are the CEPC (Circular Electron Positron Collider) in China and the FCC (Future Circular Collider) at CERN (Switzerland). To perform experiments on flavor physics in the energy range of the projects an excellent particle identification up to momenta of 30\,GeV/c is required. Several concepts of RICH detectors based on aerogel were considered and evaluated with the help of GEANT4 simulation: FARICH (Focusing Aerogel RICH) based on multilayer aerogel with maximal refractive index of 1.008, RICH with Fresnel lens based on aerogel with refractive index of 1.008, RICH based on transparent aerogel fibers with refractive index of 1.008. The results of the simulation are presented. Some results of beam tests at the BINP performed to validate GEANT4 simulation are presented as well. 
\end{abstract}
\end{frontmatter}


\section{Introduction}
\label{sec:intro}
Future e$^+$e$^-$ Higgs-factories such as the Future Circular Collider electron-positron mode (FCCee)  at CERN and the Circular Electron-Positron Collider in China (CEPC) have an extensive physics program at the energy of Z-boson production, the so called Z-pole ($\sqrt{s}=91.2$~GeV). About $4\times10^{12}$  Z-bosons  are expected to be produced at an integrated luminosity of ($\int Ldt\approx100ab^{-1}$). Such statistics will provide an extensive number of $b\overline{b}$, $c\overline{c}$,  and $\tau^+\tau^-$ pairs produced in clean Z decays for precise flavor physics investigations~\cite{fwcepc2024}.  For fruitful realization of the physics program, the reliable $\pi/K$-separation is needed to suppress combinatorial background and to separate final states with similar topologies such as:
$B_s^0\to\pi^+\pi^-$, $B_s^0\to K^+K^-$,$B_s^0\to K^{\pm}K^{\pm}$ and so on.

A baseline option of the CEPC detector is able to provide $\pi/K$-separation at the level of two standard deviations (2$\sigma$) up to 20~GeV/$c$ with the help of the combination of the $dE/dx$ and the time-of-flight ($ToF$) techniques~\cite{cepcpid2023}. In this paper, several approaches based on aerogel RICH detectors are considered for $\pi/K$-separation above 20~GeV/c at the level better than 3$\sigma$. 

\section{Aerogel with refractive index n=1.008}
\begin{figure}[htbp]
\centering
\includegraphics[width=.95\textwidth]{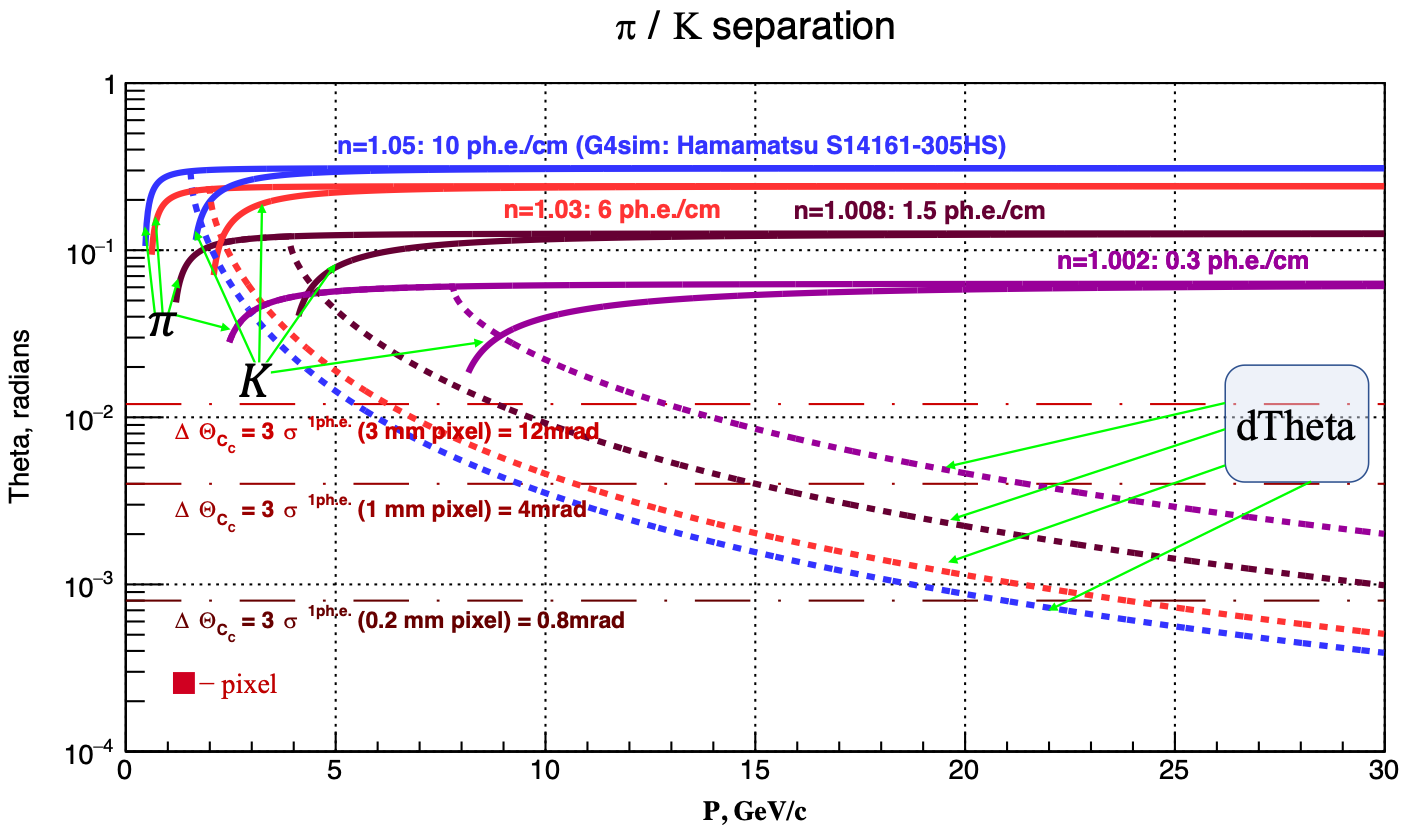}
\caption{The dependences of Cherenkov angles for pions and kaons (solid lines) and the difference between them (dashed lines) on the particle momentum are presented for the following media: aerogel with refractive index n = 1.05 (blue), aerogel with n = 1.03 (red), aerogel with n = 1.008 (brown), and $C_4F_{10}$ under pressure 5~atm with n=1.002. The number of detected Cherenkov photons per 1~cm of track length in media with different refractive indexes was recalculated from GEANT4 simulations performed for aerogel with refractive index n=1.05 and using a SiPM array from Hamamatsu S14161-3050HS as the photon detector. \label{fig:thetach}}
\end{figure}

In Fig.~\ref{fig:thetach}, the dependencies of the Cherenkov angles for pions and kaons (solid lines) and the difference between them (dashed lines) on the particle momentum are presented. Since the number of detected Cherenkov photons ($N_{ph.e.}$) depends on the refractive index and the relative velocity of the particle ($\beta=\frac{v}{c}$) as follows:
\begin{equation*}
N_{ph.e.}\sim\sin^2\Theta_C=\frac{n^2\beta^2-1}{n^2\beta^2},
\end{equation*}
the expected number of detected photons per 1~cm of track length in media with different refractive indexes was recalculated from the results of GEANT4 simulations performed for aerogel with n=1.05 and a photon detector based on SiPM arrays S14161-3050HS from Hamamatsu. These numbers are also presented in Fig.~\ref{fig:thetach}, along with the levels of three standard deviations expected for different single-photon angle resolutions which depend on the pixel sizes (3$\times$3~mm, 1$\times$1~mm and 0.2$\times$0.2~mm). It is clear from Fig.~\ref{fig:thetach} that to provide the reliable $\pi/K$-separation with the help of aerogel with n=1.008 up to a momentum P>20~GeV/$c$, it is necessary to have more than 10 detected Cherenkov photons per track in the case of a photon detector with a $\square$1$\times$1~mm pixel. This means that the aerogel thickness should be about 6$\div$8~cm. Therefore, it is necessary to have highly transparent aerogel to reduce the number of Rayleigh-scattered Cherenkov photons as well as a focusing system to reduce the impact on resolution associated with the uncertainty of the Cherenkov photon's emission point.

\begin{figure}[htbp]
\centering
\includegraphics[width=.5\textwidth]{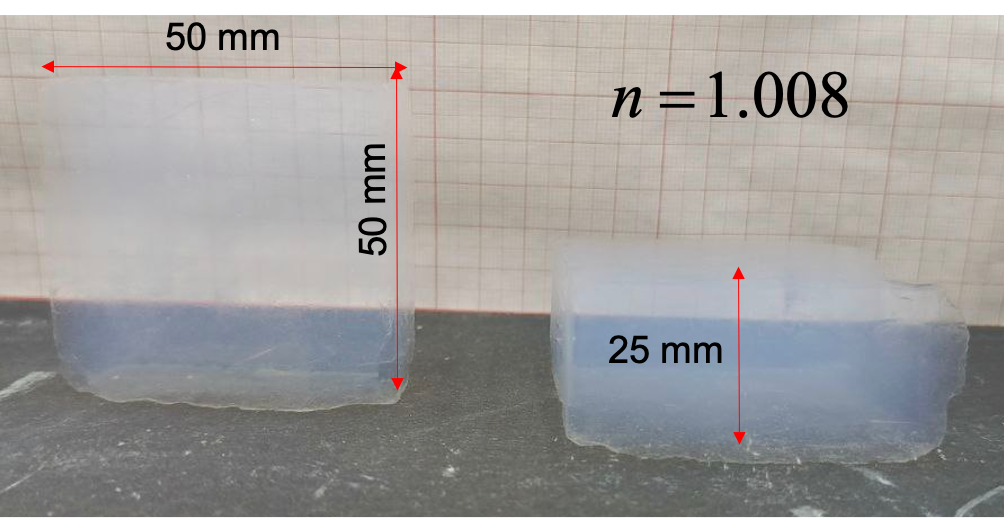}
\qquad
\includegraphics[width=.4\textwidth]{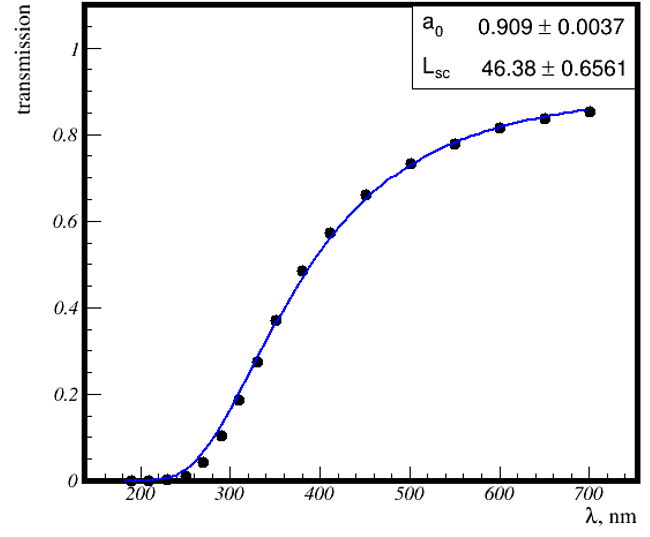}
\caption{Aerogel with a refractive index of n=1.008 produced in Novosibirsk: a picture of several blocks with their dimensions (left) and the measured transparency fitted by the Hunt formula (see eq.~\ref{eq:hunt}) with its parameters (right). \label{fig:aer1008}}
\end{figure}

In Fig.~\ref{fig:aer1008} aerogel with a refractive of n=1.008 produced in Novosibirsk is presented: a picture of several blocks with dimensions 50$\times$50$\times$25~mm (left) and the measured dependence of transparency on wavelength (right). The dependence of transparency ($T$) on wavelength ($\lambda$) is fitted by the Hunt formula:
\begin{equation}
T=\frac{I}{I_0}=A_0e^{-\frac{d}{L_{sc}\cdot\left(\lambda/400\right)^4}},
\label{eq:hunt}
\end{equation}
where $A_0$ is a fit parameter corresponding to reflection and scattering of light at the aerogel surfaces, $d$ is the thickness of the aerogel and $L_{sc}$ is the fit parameter equal to the Rayleigh light scattering length at a wavelength of 400~nm. For the presented samples, $A_0$=0.91 and $L_{sc}$=4.6~cm.

\begin{figure}[htbp]
\centering
\includegraphics[width=.45\textwidth]{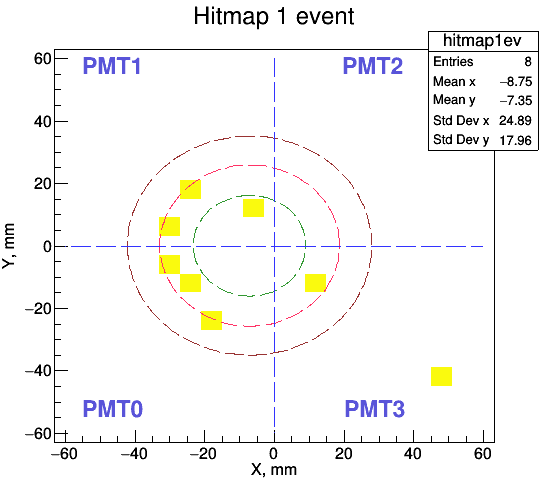}
\qquad
\includegraphics[width=.45\textwidth]{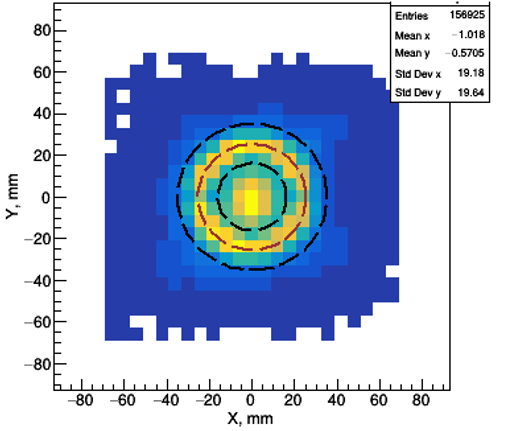}
\caption{Cherenkov hit maps detected at the BINP beam test with relativistic electrons (2.5\,GeV) and aerogel with n=1.008: for a single track (left) and accumulated for 13\,000 tracks (right). \label{fig:hitmap}}
\end{figure}

Several such aerogel tiles were tested with relativistic electrons (2.5~GeV) at the Budker Institute of Nuclear Physics (BINP) beam test facilities~\cite{exbeam2016} to investigate how many Cherenkov photons per track could be detected in the ring of a RICH counter based on aerogel with n=1.008. Some details on the beam line equipment can be found in~\cite{farich2020}. To detect Cherenkov radiation, four multi-anodes PMTs (MaPMT) H12700 (Hamamatsu), each with an 8$\times$8 pixel array of  6$\times$6\,mm pixels, were used. The radius of the Cherenkov ring from aerogel with a refractive index of n=1.008 at the distance 160\,mm will be approximately 20\,mm and the Cherenkov ring will  fit into a single MaPMT with an active area of 46$\times$46\,mm.  The hit maps of photons produced by electron tracks in aerogel with a thickness of 50\,mm (a stack of two tiles, 25+25\,mm thick) are shown in Fig.~\ref{fig:hitmap}: for a single track (left) and for thousands of accumulated tracks (right). To obtain the right picture in Fig.~\ref{fig:hitmap}, the center of each ring was reconstructed by fitting the Cherenkov hits distribution to a circle, and then the center of each event was subtracted.

Three thick aerogel radiators were tested under the same conditions: 25\,mm, 25+25=50\,mm and 25+25+25=75\,mm. A summary of the beam test results is given in the Table~\ref{tab:tbres}. In the table $N_{pe}$ is the number of detected Cherenkov photons and $\sigma_R^{SPE}$ is the Single PhotoElectron (SPE) resolution, which was obtained from the distribution of each detected photon in radius ($R^{SPE}$), evaluated as the difference between the hit position ($X_{hit},Y_{hit}$) and the position of the ring center ($X_{0},Y_{0}$) reconstructed for each event:
\begin{equation*}
R^{SPE}=\sqrt{(X_{hit}-X_0)^2+(Y_{hit}-Y_0)^2}.
\end{equation*}
The value of $\sigma_R^{SPE}$ for the aerogel sample with a thickness 25\,mm is not presented in the table because, in this case, when the average number of detected hits is about 3, it could be calculated for only half of the events and would not be a representative value.
\begin{table}[htbp]
\centering
\caption{Results of the beam tests performed with relativistic electrons and aerogel n=1.008 at the BINP beam test facilities.\label{tab:tbres}}
\smallskip
\begin{tabular}{|l|c|c|c|}
\hline
&25\,mm& 25+25\,mm & 25+25+25\,mm\\
\hline
$N_{pe}$ & $3.8\pm0.05$ & $7. 7\pm0.05$ & $8.7\pm0.05$\\
\hline
$\sigma_R^{SPE}$, mm & -- & $2.94\pm0.03$  & $3.37\pm0.07$\\
\hline
$\sigma_{aer}$, mm & -- & $2.37\pm0.03$  & $2.89\pm0.07$\\
\hline
$\sigma_{\Theta_c}^{SPE}$, mrad & -- & $14\pm0.2$  & $18\pm0.4$\\
\hline
\end{tabular}
\end{table}
There is an impact of pixel size in the single photon resolution, which can be estimated as $6/\sqrt{12}=0.87$\,mm and can be subtracted in quadrature from the experimental data to obtain the contribution to the Cherenkov ring resolution coming only from the aerogel ($\sigma_{aer}$). Taking into account the distance from the aerogel to the photon detector plane (160\,mm), the Cherenkov angle resolution can be expressed in milliradians as ($\sigma_{\Theta_c}^{SPE}$ (see Table~\ref{tab:tbres}). From the results presented in the table, it is possible to conclude that the optimal aerogel thickness with n = 1.008 is about 6\,cm. To increase the number of detected photons per track, it is necessary to improve the aerogel transparency (increase the Rayleigh light scattering length) and the photon detection efficiency (PDE) of the photon detectors. For instance, in these tests, MaPMT with about 20\% quantum efficiency at maximum (400\,nm) were used, while it is possible to use other photon detectors, for example, those based on Silicon Photomultiplayer (SiPM) arrays like S14161-3050HS (Hamamatsu), with a PDE of about 45$\div$50\% at maximum (450\,nm). An increase in the number of detected photons will improve overall resolution per track as $\sim\frac{1}{\sqrt{N_{pe}}}$), but even an increase of $N_{pe}$ by a factor of two will not be enough to compensate for the impact of the aerogel thickness (uncertainty of the emission point). Therefore, several approaches with proximity focusing were investigated with the help of GEANT4 simulation. The main results of the simulations are presented in the following sections.
\section{FARICH based on multilayer aerogel with a maximal refractive index of \texorpdfstring{$n_{\text{max}}=1.008$}{}}
\label{sec:farich}
\begin{figure}[htbp]
\centering
\includegraphics[width=.3\textwidth]{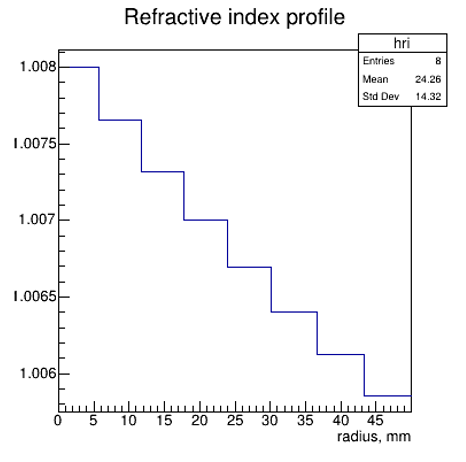}
\qquad
\includegraphics[width=.6\textwidth]{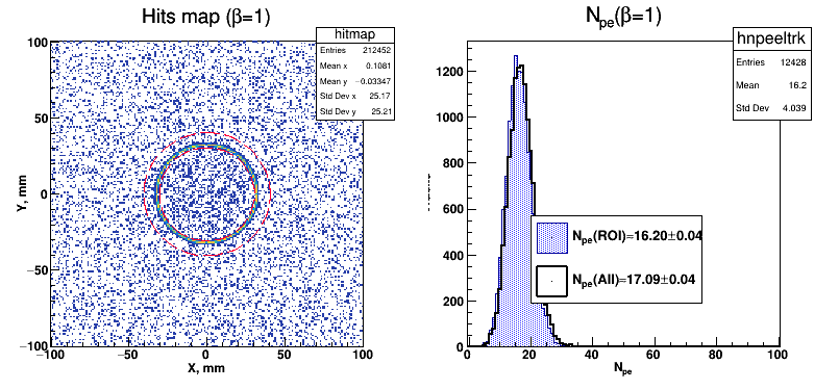}
\caption{Results of a GEANT4 simulation for a FARICH based on an 8-layer focusing aerogel: the refractive index profile (left), a Cherenkov photon hit map (center), and the number of detected photons per track (right). \label{fig:farichg4}}
\end{figure}

The FARICH (Focusing Aerogel RICH) technique was first proposed in 2004~\cite{farich2005,arich2005}. Recent progress on FARICH R\&D can be found in~\cite{farich2024}. To provide reliable $\pi/K$-separation for momenta above 20\,GeV/c, an 8-layer focusing aerogel with a maximal refractive index of $n_{max}$=1.008 and a total thickness of 50\,mm was implemented in the simulation. The photon detector was represented by an array of SiPMs with a pixel size of 1$\times$1\,mm and a PDE matching that of the SiPM S14161-3050HS (Hamamatsu). The optical parameters measured for the aerogel presented in Fig.~\ref{fig:aer1008} were used in the simulation. The focal length (the distance between the input surface of the aerogel and the plane of the photon detector) is  250\,mm.
In Fig.~\ref{fig:farichg4}, some results of the GEANT4 simulation for the FARICH option based on an 8-layer focusing aerogel are presented: the refractive index profile (left), a Cherenkov photon hit map (center) and the number of detected photons per track (right). 

It is shown with the help of simulations that a Cherenkov angle resolution of about 0.3\,mrad per track is achievable with such an approach, and it would be sufficient to provide $\pi/K$-separation at a momentum of 30\,GeV/c, where the difference in Cherenkov angles is about 1\,mrad.

\begin{figure}[htbp]
\centering
\includegraphics[width=.45\textwidth]{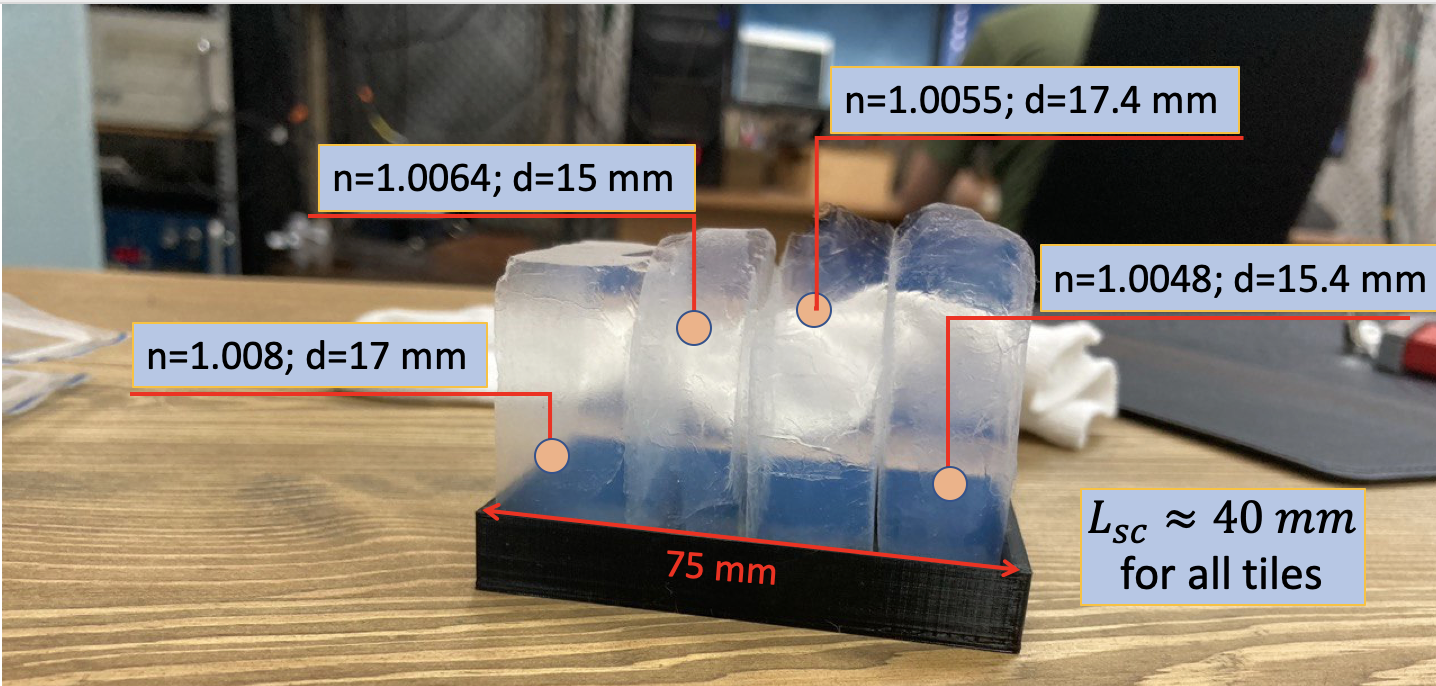}
\qquad
\includegraphics[width=.45\textwidth]{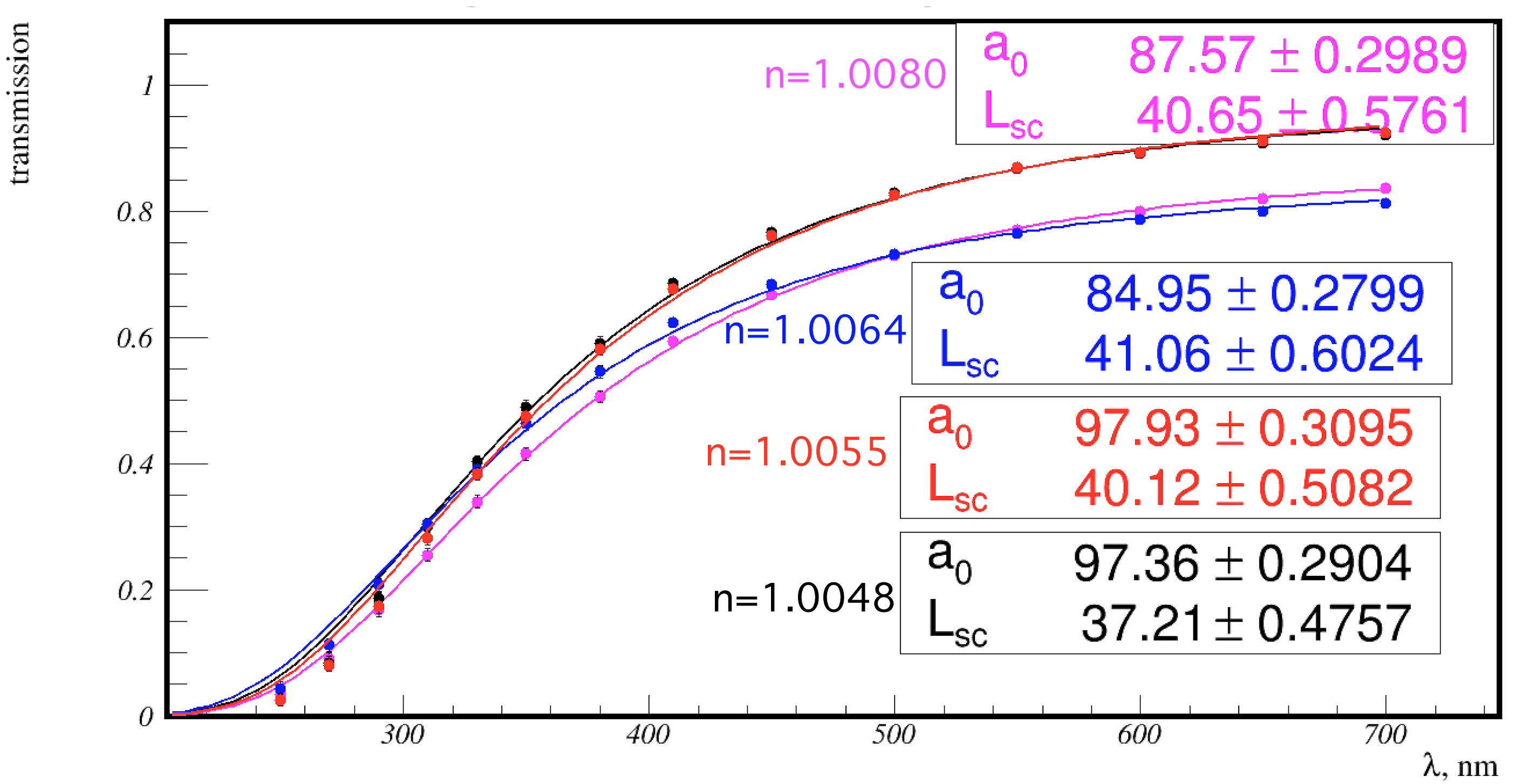}
\caption{Ultralight aerogel produced in Novosibirsk: a picture of a prototype FARICH radiator consisting of several blocks with their thicknesses and refractive indices (left) and  measured transparency fitted by the Hunt formula (see eq.~\ref{eq:hunt}) with its parameters (right). \label{fig:aerless1008}}
\end{figure}

To test the FARICH option based on ultralight aerogel, several ultralight aerogel tiles with large light scattering lengths (approximately 40\,mm) were produced in 2025. A 75\,mm thick FARICH radiator was assembled from these samples (see Fig.~\ref{fig:aerless1008}) to be tested with relativistic electrons at the Budker Institute of Nuclear Physics (BINP) beam test facilities~\cite{exbeam2016} for the first time in the near future.

\section{RICH with a Fresnel lens based on aerogel with a refractive index of \(n=1.008\)}
\label{sec:mrich}
\begin{figure}[htbp]
\centering
\includegraphics[width=.35\textwidth]{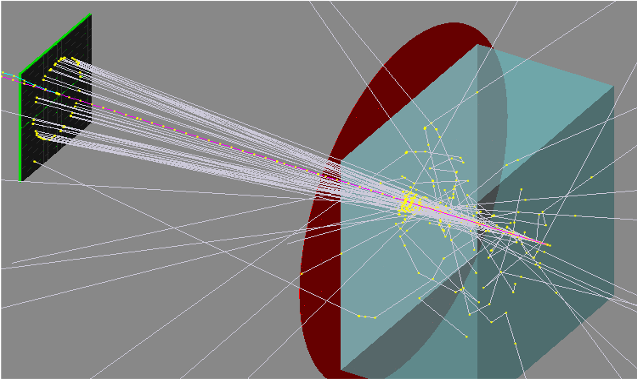}
\qquad
\includegraphics[width=.55\textwidth]{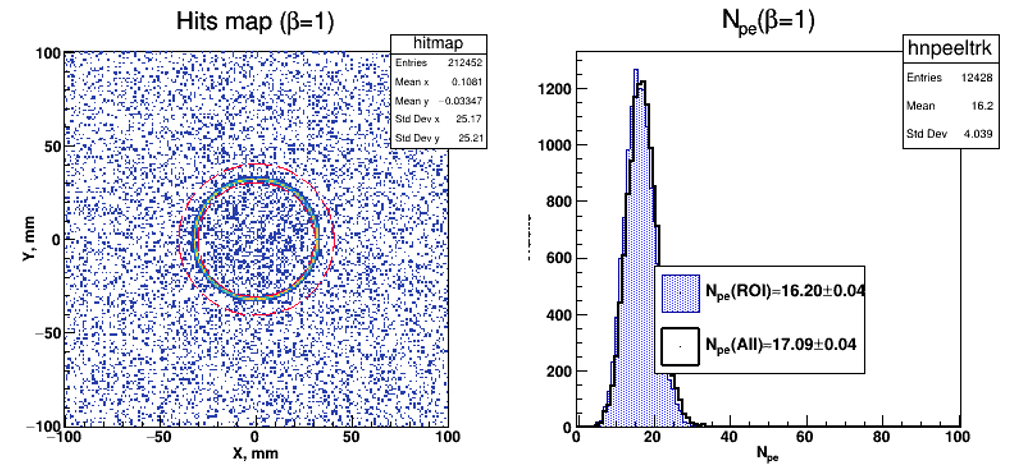}
\caption{Results of a GEANT4 simulation for the mRICH approach based on aerogel with n=1.008 and a thin acrylic Fresnel Lens: an illustration of the scheme implemented in the GEANT4 simulation (left), a Cherenkov photon hit map (center), and the number of detected photons per track (right). \label{fig:mrichg4}}
\end{figure}

Another approach for focusing of Cherenkov light from the aerogel was inspired by recent progress in the R\&D of a modular RICH for the EIC experiment~\cite{mrich2024}, where Cherenkov light from a thick aerogel radiator is focused by means of a thin acrylic Fresnel lens. For the simulation, aerogel with n=1.008, a thickness 6\,cm, and the same transparency as presented above was used. For the photon detector, the same parameters as those described in sec.~\ref{sec:farich} were used. The optical parameters of the Fresnel Lens were taken from~\cite{ultra2007}, where Fresnel Lens were used in the ULTRA experiment to collect Cherenkov light produced by extensive air showers. The focal distance of the Fresnel lens in our simulation is 10 inches. In Fig.~\ref{fig:mrichg4}, some results from the simulation are presented: an illustration of the scheme implemented in the GEANT4 simulation (left), a Cherenkov photon hit map (center), and the number of detected photons per track (right). 

It is shown that such an approach is also able to provide Cherenkov angle resolution sufficiently small  to ensure reliable $\pi/K$-separation at a momentum of 30\,GeV/c.

\section{RICH based on aerogel fibers with a refractive index of \(n=1.008\)}
\begin{figure}[htbp]
\centering
\includegraphics[width=.35\textwidth]{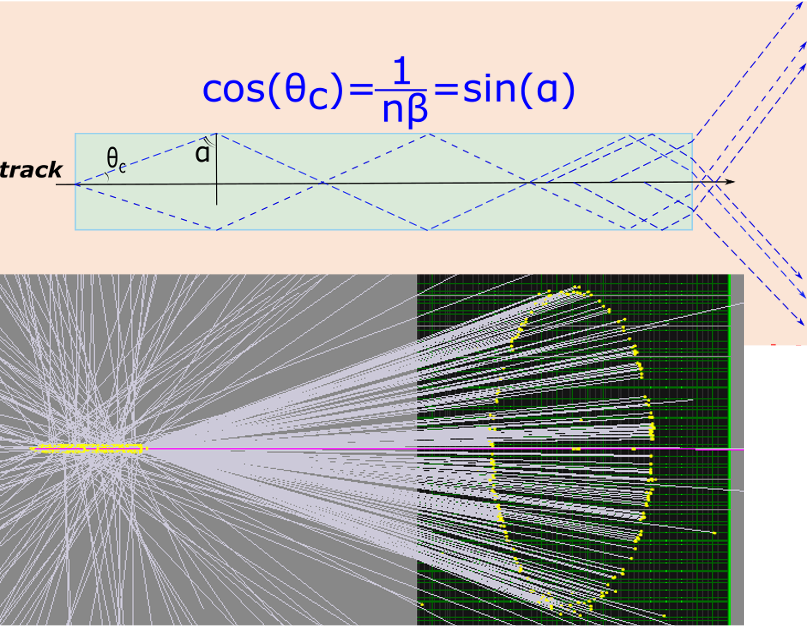}
\qquad
\includegraphics[width=.55\textwidth]{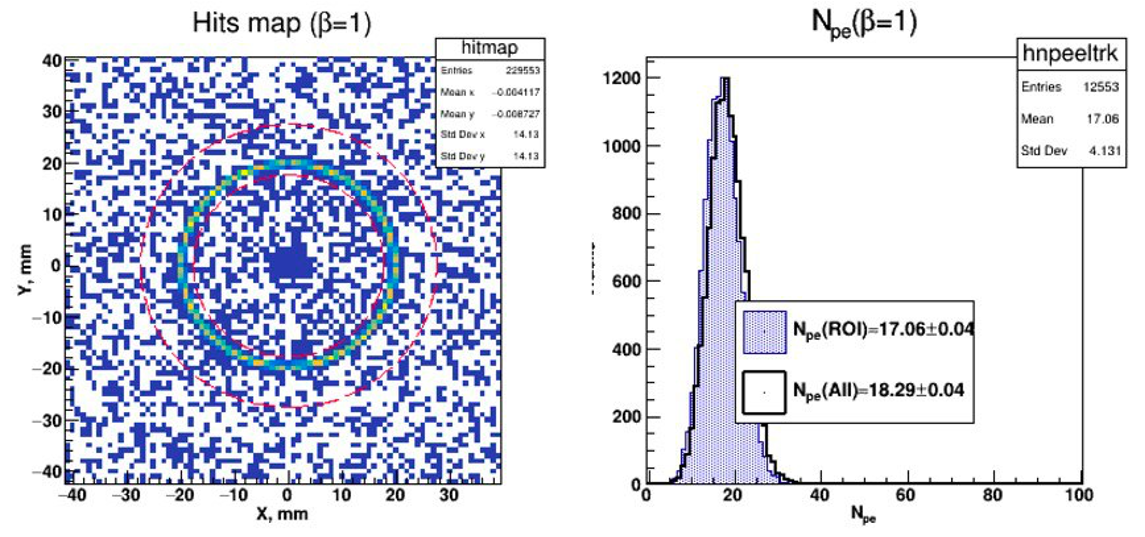}
\caption{Results of a GEANT4 simulation for the RICH approach based on aerogel fibers with n=1.008: an illustration of the concept and the scheme implemented in the GEANT4 simulation (left), a Cherenkov photon hit map (center), and the number of detected photons per track (right). \label{fig:firichg4}}
\end{figure}

Recent progress in the production of transparent aerogel fibers, as presented in~\cite{fiberaer2023}, allows us to consider another option for a RICH detector based on aerogel fibers. It turns out  that Cherenkov light produced by a charged particle traveling along the fiber will be trapped under total internal reflection, as shown in Fig.~\ref{fig:firichg4}(left). In this case, the uncertainty of the Cherenkov emission point will be determined by the diameter of the fiber and will not depend on the fiber length. The optical  parameters of the aerogel used in the simulation were taken to be the same as for the previous options: the fiber length is 6\,cm and the diameter is 0.4\,mm. The photon detector parameters were taken to be the same as those in described in  Sec.~\ref{sec:farich}. In Fig.~\ref{fig:firichg4}, the results of the simulation for the RICH based on aerogel fibers are presented: an illustration of the geometry description in GEANT4 and the tracking of a photon for one simulated track (left, bottom), a hit map for detected Cherenkov photons (center), and the distribution of events as a function of the number of detected Cherenkov photons (right).

\section{On photon detection options with sub-millimeter spatial resolution}
One of the major assumptions is that the position of the Cherenkov photons will be determined with a spatial resolution on the level of 200\(\div\)300\(\mu\)m. Such spatial resolution will impact the single-photon Cherenkov angle resolution by about 1 mrad for a focusing distance of 300\,mm. The impact from aerogel refractive index dispersion can also be estimated at about 0.7\,mrad. The estimation is based on simulation results and the empirical equation from~\cite{aerdisp} that connects refractive index dispersion with the bulk density of aerogel. This means that the impact from the photon detector spatial resolution is approximately equal to the impact from refractive index dispersion of aerogel with a refractive index of 1.008. Both effects will decrease with the number of detected photons (\(N_{hit}\)) as $1/\sqrt{N_{hit}}$, and for N\(_{hit}\)=10, the impact from both effects will be about 0.5\,mrad, which should be sufficient to provide \(\pi/K\)-separation up to 25\,GeV/c (see Fig.~\ref{fig:thetach}). 

To provide a spatial resolution of about 300\,\(\mu\)m, the pixel size should be \(\leq\)1\,mm. There are several photon detector types that fit these requirements:
\begin{itemize}
    \item[--] Microchannel plate (MCP) based PMTs, such as the XP85122 from Photonis (USA) with an active area of 53\(\times\)53\,mm and an anode pitch of 1.6\,mm, or the MAPT253 from Photek Ltd. (UK) with a similar active area and a 64\(\times\)64 pixel array with an anode pitch of 0.828\,mm;
    \item[--] SiPM arrays, such as the JARY-TN1050-8\(\times\)8C 64-channel array from Joinbon Technology Co., Ltd. (China) with a 1\(\times\)1\,mm active sensor area, or the HPK S13552 128-channel linear arrays from Hamamatsu (Japan) having a 0.23\(\times\)1.625\,mm\(^2\) active area packaged with a 0.25\,mm linear pitch, among others.
\end{itemize}

For all these options, the issue of readout electronics is very crucial, but it is being solved for several projects with the help of the development of dedicated ASIC (Application Specialized Integrated Circuit) chips and DAQ (Data Acquisition) systems. As an example, the R\&D for the TORCH detector of the LHCb experiment could be considered~\cite{torch2022}. From the point of view of minimizing the number of readout electronics channels, it is  interesting to use SiPMs with a resistive sensitive anode, such as LG-SiPMs (linear-grade Silicone PhotoMultipliers) from FBK (Italy)~\cite{lgfbk2020} or PSS-SiPM (position-sensitive SiPMs) from NDL (China), as well as other similar products from various manufacturers. In such devices, all the micro APD (avalanche photo diode) cells are connected to the a cap resistive layer (CRL) to implement a charge division mechanism. The analog signals are readout by 4 measurers from the 4 edges of the array, and the exact position could be restored through the ratio of charge shared between the channels. The precision of such hit determination is declared to be as small as 200\,\(\mu\)m in the case of a single photon event. Such an approach allows us to use only 4 readout electronics channels to reconstruct the photon position from 3\(\times\)3\,mm or 6\(\times\)6\,mm sensors with high spatial resolution. The main advatages of this technique are a reduction in the number of electronics channels and consequently lower power consumption near the silicone sensors, which are usually very sensitive to the ambient temperature. The disadvantages are the affects on photon position reconstruction from intrinsic noise and from simultaneous hits  of two or more photons in one sensor. To suppress the first affect, an effective cooling system is required, while for the second effect, geometry optimization for the task and experimental condition is needed.

\section{Summary and discussion of results}

The GEANT4 simulation performed for aerogel with refractive index of n=1.008 produced in Novosibirsk was validated by the results of the beam test with relativistic electrons at the BINP beam test facilities. The results show us that it is possible to build a RICH detector based on aerogel with n=1.008 and modern photon detectors, in which the average number of detected photons from a relativistic particle will be about 16.

\begin{figure}[htbp]
\centering
\includegraphics[width=.36\textwidth]{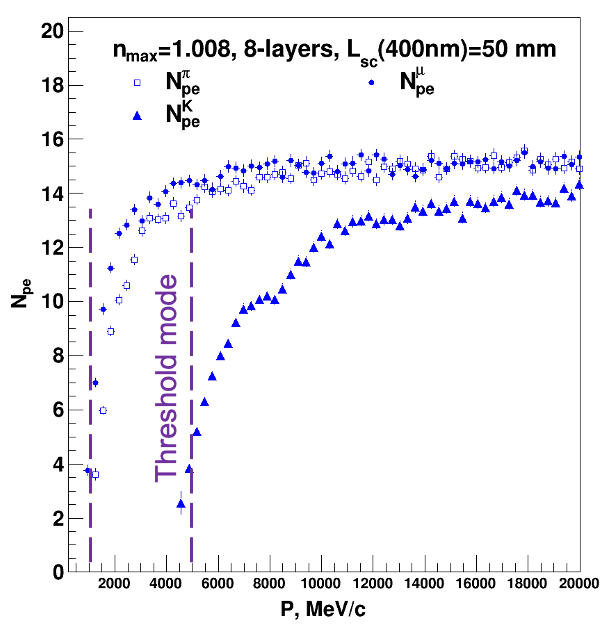}
\qquad
\includegraphics[width=.55\textwidth]{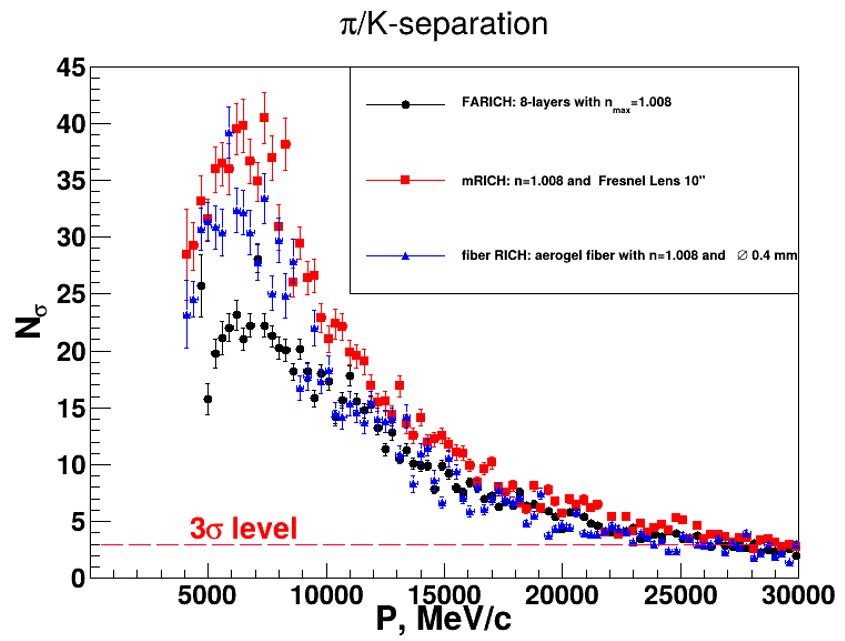}
\caption{Dependence of the number of detected Cherenkov photons (left) and the quality of $\pi/K$-separation in terms of the number of $\sigma$ (right) on the particle momentum for three approaches based on aerogel with n=1.008: FARICH with an 8-layer focusing aerogel (black filled circles), RICH based on aerogel fibers (blue filled triangles), and aerogel RICH with a thin acrylic Fresnel lens (red filled squares). \label{fig:piksep}}
\end{figure}

Three approaches for proximity-focused RICH detectors based on aerogel with n=1.008 were investigated with the help of GEANT4 simulations.
All three approaches considered demonstrate very promising results. In Fig.~\ref{fig:piksep}, a comparison of the three approaches is presented: the dependence of the number of detected Cherenkov photons (left) and the quality of separation in terms of the number of standard deviations ($\sigma$) (right) on the particle momentum are shown. It is shown that all three approaches are capable of providing $\pi/K$ -separation from 5 to 30\,GeV/c at a level better than 3$\sigma$, while in the momentum range of 1.5 to 5\,GeV/c, it is possible to separate pions and kaons in threshold mode.

The possibility of producing highly transparent aerogels (Rayleigh light scattering length $L_{sc}(400nm)\geq40mm$)  with refractive indices below or equal to 1.008 (1.0048, 1.0055, 1.0064) has been demonstrated.

For all considered options, the key issues are the photon detector with precise position sensitivity and readout electronics with a high number and density of channels. Therefore, position-sensitive SiPMs such as the PSS 11-3030-S from NDL (China) or the linear-grade SiPM from FBK (Italy)~\cite{lgfbk2020} are very attractive options for providing precise position sensitivity with high PDE (Photon Detection Efficiency) and a moderate number of readout channels. For instance, the position of detected photons in such devices is reconstructed with the help of charge shared between 4 pads for each square 6$\times$6=36\,mm$^2$ pixels, and the accuracy of such reconstruction can be as small as 0.2\,mm according to the datasheet and results presented in~\cite{lgfbk2020}.

\end{document}